\documentclass[preprint,12pt]{elsarticle}

\usepackage{color}

\usepackage{amssymb}

\journal{Foundations of Computational Mathematics}

\begin{document}

\begin{frontmatter}



\title{A new topological entropy-based approach for measuring similarities among piecewise linear functions}

\author[1,3,$^*$]{Matteo Rucco}
\author[2,$^*$]{Rocio Gonzalez-Diaz}
\author[2]{Maria-Jose Jimenez}
\author[2]{Nieves Atienza}
\author[3]{Cristina Cristalli}
\author[3]{Enrico Concettoni}
\author[3]{Andrea Ferrante}
\author[1]{Emanuela Merelli}

\address[1]{University of Camerino, School of Science and Technology, Computer Science Division, Camerino, IT}
\address[2]{University of Seville, School of Computer Engineering, Applied Math Dept,  Seville, Spain}
\address[3]{Research for Innovation Group. Loccioni Group, Angeli di Rosora (AN), IT\\
$^*$Corresponding E-mails: matteo.rucco@unicam.it, rogodi@us.es}


\begin{abstract}
In this paper we present a novel methodology based on a topological entropy, the so-called persistent entropy, for addressing the comparison between discrete piecewise  linear functions. The comparison is certified by the {\it stability theorem for persistent entropy}. The theorem is used in the implementation of a new algorithm. The algorithm transforms a discrete piecewise linear function into a filtered simplicial complex that is analyzed with persistent homology and persistent entropy. 
Persistent entropy is used as discriminant feature for solving the supervised classification problem of real long length noisy signals of DC electrical motors. The quality of classification is stated in terms of the area under receiver operating characteristic curve (AUC=94.52\%).
\end{abstract}


\end{frontmatter}


\section{Introduction}
\label{sec:intro}
Piecewise linear function (PL) is a powerful mathematical tool largely used for approximating signals. The task of measuring the similarity among piecewise linear functions (PLs) is still an open issue and a solution is strongly required in machine learning methods. The comparison between the area under the curves (AUCs) of discrete digital signals is a weak measure: for each value of AUC a family of infinite signals exists. Several approaches for measuring the similarities among PLs have been reported in literature, at the best of our knowledge the most relevant techniques are:
\begin{enumerate}
\item distance-based methods, e.g. Pompeiu-Hausdorff distance~\cite{huttenlocher1993comparing};
\item similarities based on global descriptors: time-frequency analysis, wavelet analysis, mutual information, (Normalized/Zero-Mean) cross-correlation, sum of squared differences, etc\dots~\cite{cohen1995time,viola1997alignment};
\item distances among ``bags of local features''~\cite{delaitre2010recognizing}.
\end{enumerate}
Generally, the identification of common patterns among signals suffer of the shifting problem~\cite{papoulis1977signal}. Formally, given two 2-dimensional signals $(S_1,S_2)$ - that are two ordered collection of real points - the signal $S_2$ is shifted respect to signal $S_1$ if $(x_2=x_1+n~and/or~y_2=y_1+m$,where $m,n \in \mathbb{R})$. For this reason, the distance-based methods without pre-alignments can not be directly applied. Pre-alignments techniques need in general high time-consuming. The potential solution is represented by the dynamic time warping (DTW)~\cite{salvador2004fastdtw}. However, DTW is not computationally convenient for compiling a pair-wise distance matrix in case of several long-length signals~\cite{keogh2005exact}. {\it Similarities based on global descriptors} and {\it distances among ``bags of local features''} both need the extraction of local or global features from the PLs. The features can be used for extrapolating several useful information (e.g., periodicity, state transitions, chaotic behaviors, etc\dots) but there is not a unique criteria for deciding which features are completely informative and often it is mandatory to execute a feature selection (or reduction) procedure.

In this paper we present a new method that executes the comparison among the shape of PLs. The methodology described in this article is based on topology. Roughly speaking, topology is a branch of pure mathematics that deals with the analysis of the shapes. Briefly, our contribution represents a PL by a topological space, i.e. a filtered simplicial complex, that is qualitatively described by persistent homology and quantitatively measured by the persistent entropy. Thanks to the {\it stability theorem for persistent entropy}, persistent entropy is used as unique global feature for comparing signals. We observe that this quantity is invariant respect to the shifting of the two coordinates. It means, given two signals with the same shape, even if they are shifted along one or both directions they have the same persistent entropy. The paper is organized as follows, in Section~\ref{sec:back} we remark the mathematical background for understanding the methodology described in Section~\ref{sec:method}. Section~\ref{sec:method} also contains either the proof for the {\it stability theorem for persistent entropy} and an algorithm for the comparison between PLs. The application of the methodology to a real case study is reported in Section~\ref{sec:results}. Section \ref{sec:conclusions} is devoted to briefly remark the main results of our paper and relevant observations are pinpointed out.

\section{Background}
\label{sec:back}
\subsection{Topology}
A topological space is a powerful mathematical concept for describing the connectivity of a space. Informally, \textit{a topological space is a set of points each of them equipped with the notion of neighboring}. More formally, it is defined by equipping a set with a \textsc{topology} as follows:

\vspace{0.5cm}\noindent \textbf{Topology.}
A topology on a set of point $\mathbb{X}$ is a subset $T \subseteq 2^\mathbb{X}$
\begin{itemize}
\item[a)] If $S_1, S_2 \in T$, then $S_1  \cap S_2 \in T$;
\item[b)] If $\{S_j | j \in J\} \subseteq T$, then $\bigcup_{j\in J}S_j \in T$;
\item[c)] $\emptyset, \mathbb{X} \in T$.
\end{itemize}

\vspace{0.5cm}
\noindent \textbf{Topological space.}
The pair $(\mathbb{X},T)$ of a set $\mathbb{X}$ and a topology T is a topological space $\mathfrak{C}$.
\vspace{0.5cm}

One way to represent a topological space is by decomposing it into {\it simple} pieces such that their common intersections are lower-dimensional pieces of the same kind. In this paper, we use (abstract) simplicial complexes as the data structure to represent topological spaces.

\vspace{0.5cm}
\noindent {\bf Abstract simplicial complex.} An  abstract simplicial complex ${\cal K}$ is given by:
\begin{itemize}
\item a set $V$ of $0-$simplices;
\item for each $k\geq 1$ a set of $k-$simplices $\{\sigma=\{v_0, v_1, \dots, v_k\}$, where $v_i \in V\}$;
\item each $k-$simplex has $k+1$ faces obtained removing one of the vertices;
\item if $\sigma$ belongs to ${\cal K}$, then all faces of $\sigma$ must belong to ${\cal K}$.
\end{itemize}

\vspace{0.5cm}
\noindent {\bf Simplicial complex.} 
A  simplicial complex $K$ is a geometrical realization of an abstract simplicial complex $\cal{K}$.
A simplicial complex is obtained by a nested family of simplices: a $0-$simplex can be thought as a point, a $1-$simplex as  an edge, a $2-$simplex as a filled triangular face and a $3-$simplex as a filled tetrahedron.  
\vspace{0.5cm}

See \cite{hatcher2002algebraic} and \cite{munkres1984elements} for an introduction to algebraic topology.

\subsection{Persistent homology}
Homology is an algebraic machinery used for describing a topological space $\mathfrak{C}$. The $k-$Betti number represents the rank of the $k-$dimensional homology group.
Informally, for a fixed $k$,
the $k-$\textit{Betti number} counts
 the number of $k-$dimensional holes characterizing $\mathfrak{C}$:
$\beta_0$ is the number of connected components, $\beta_1$ counts the number of
holes in 2D or tunnels in 3D\footnote{nD refers to the $n-$dimensional space $\mathbb{R}^n$.}, $\beta_2$ can be thought as the number of voids in geometric solids.
Persistent homology  is a method for computing $k-$dimensional holes at different spatial resolutions. 
Persistent holes 
are more likely to represent true features of the underlying space, rather than artifacts of sampling (noise), or particular choice of parameters.
For a more formal description we refer to~\cite{edelsbrunner2010computational}.  

In order to compute persistent homology, we need a distance function on the underlying space. This can be obtained constructing \textbf{a filtration of the simplicial complex},  that is a nested sequence of increasing subsets. More formally, a filtered simplicial complex $K$ is a collection of subcomplexes
$\{K(t): t \in \mathbb{R}\}$ of $K$ such that $K(t) \subset K(s)$ for $t< s$ and there exists $t_{max}\in \mathbb{R}$ such that $K_{t_{max}}=K$. The filtration time (or filter value) of a simplex $\sigma \in K$ is the smallest $t$ such that $\sigma \in K(t)$.

\vspace{0.5cm}
\noindent \textbf{Persistent homology} describes how the homology of $K$ changes along filtration. A $k-$dimensional Betti interval, with endpoints $[t_{start}, t_{end}),$ corresponds to a $k-$dimensional hole that appears at filtration time $t_{start}$ and remains until time $t_{end}$. We refer to the holes that are still present at $t= t_{max}$ as \textit{persistent topological features}, otherwise
they are considered \textit{topological noise}~\cite{adams2011javaplex}.
The set of intervals representing birth and death times of homology classes is called the {\it persistence barcode} associated to the corresponding filtration. 
Instead of bars, we sometimes draw points in the plane such that a point $(x,y)\in \mathbb{R}^2$ (with $x< y$) corresponds to a bar $[x, y)$ in the barcode. This set of points is called  {\it persistence diagram}. 

\subsection{Persistent Entropy}
In order to measure how much is ordered the construction of a filtered simplicial complex, a new entropy measure, the so-called \textit{the persistent entropy}, has been defined in~\cite{rucco2015characterisation}. A precursor of this definition was given in~\cite{chintakunta2015entropy} to measure how different bars of the barcode are in length. Here  we recall the definition.

\vspace{0.5cm}
\noindent \textbf{Persistent entropy.}
Given a filtered simplicial complex $\{K(t) :  t\in F\}$, and the corresponding persistence barcode $B = \{a_i=[x_i , y_i) : i\in  I\}$, the \textit{persistent entropy} $H$ of the filtered simplicial complex is calculated as follows:
$$
H=-\sum_{i \in I} p_i  log(p_i)
$$
where $p_i=\frac{\ell_i}{L}$, $\ell_i=y_i - x_i$, and $L=\sum_{i\in I}\ell_i$.
Note that, when topological noise is present, for each dimension of the persistence barcode, there can be more than one interval, denoted by $[x_i~,~y_i)$, with $i \in I$. This is equivalent to say that, in the persistent diagram, the point $[x_i,y_i)$ could have multiplicity greater than $1$ (see \cite[page 152]{edelsbrunner2010computational}). 
In the case of an interval with no death time, $[x_i~,~\infty)$, the corresponding barcode $[x_i~,~m)$ will be considered, where $m = \max\{F\} + 1$.

Note that the maximum persistent entropy corresponds to the situation in which all the intervals in the barcode are of equal length. In that case, $H=\log n$ if $n$ is the number of elements of $I$. Conversely, the value of the persistent entropy decreases as more intervals of different length are present.

\section{Topological comparison of plots}
\label{sec:method}
The aim of our paper is to address the problem of the comparison between the shape of plots. In order to satisfy this task, instead of using metric spaces (e.g., DTW), we propose to study the shape of the plots by topology. The methodology is completely based on algebraic topology, it transforms a plot into a filtered 1-dimensional simplicial complex that is analyzed by persistent homology. From the homological groups we  compute the persistent entropy. Persistent entropy is used as the feature that characterizes the signal. 

This section is organized as follows: in Subsection~\ref{sec:algorithm} we introduce a formal description of the methodology and we derive an algorithm for its computation; and in Subsection \ref{sec:results} we apply the methdology to a real case study.

\begin{figure}
\centering
\includegraphics[width=\textwidth]{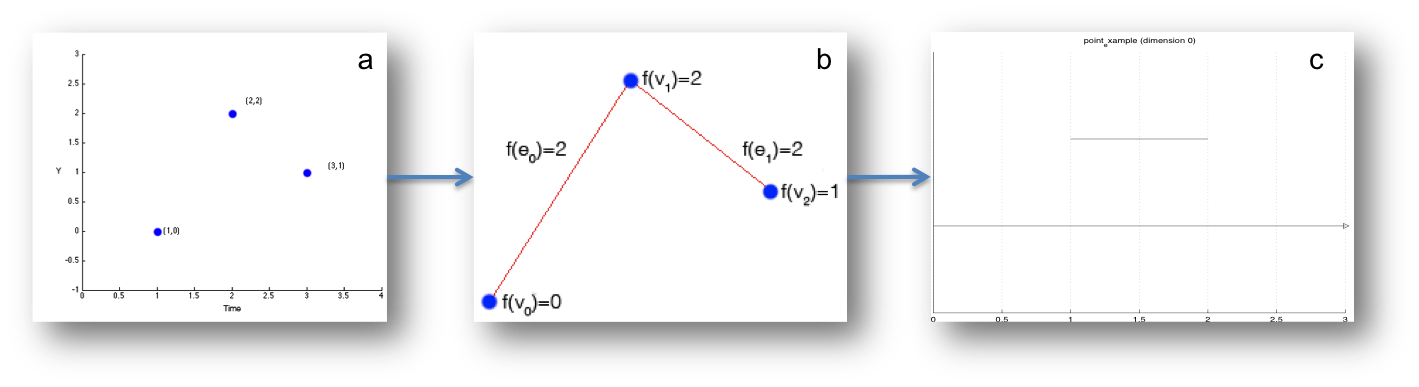}\\
\caption{Graphical representation of the methodology. See Subsection~\ref{sec:algorithm} for a complete explanation.}
\label{fig:construction}
\end{figure}

\subsection{Methodology}
\label{sec:algorithm}

Given an unknown  continuous signal $\tilde{f}:\mathbb{R}^n \to \mathbb{R}$, suppose that our input is the value of $\tilde{f}$ on  a finite set of points $S\subset \mathbb{R}^n$.

\begin{itemize}
\item Let $K$ be a simplicial complex with real values specified at all vertices in $S$.
E.g., if $S\subset \mathbb{R}$, then $K$ is a line subdivided in segments with endpoints in $D$.

\item Using linear extension over the cells of $K$, we obtain a piecewise linear $(PL)$ function $f : K \to {\mathbb R}$ (being $f(u)=\tilde{f}(u)$ for  $u\in S$).
It is convenient
to assume that $f$ is {\it generic} by which we mean that the vertices have distinct
function values. 
To ensure unique values, $f$  may need to be perturbed. 
One way of doing this is to add a
linear ramp to $f$ (see \cite[page 1650]{vanessa}).
We  can then order the vertices by increasing function value
as $f(u_1) < f(u_2) < \dots < f(u_n)$.

\item Now,  the {\it lower star} of $u_i$ can be computed which  is the subset of
simplices for which $u_i$ is the vertex with maximum function value,
$$St\_ \,u_i = \{\sigma\in St\,u_i : x\in \sigma \Rightarrow f(x) \leq f(u_i)\}.$$
The considered filtration is the {\it lower
star filtration} of $f$ (see \cite[page 135]{edelsbrunner2010computational}): $\emptyset=K_0\subset  K_1\subset \cdots \subset K_n = K$,
in which $K_i$ is the union of the first $i$ lower stars.

\item Finally,  persistent entropy $H(f)$ is computed.
\end{itemize}


\noindent{\bf Persistent Entropy Stability Theorem.}
Given two PL functions on simplicial complexes embedded in $\mathbb{R}^n$, $f : K \to {\mathbb R}$ and  $g : K' \to {\mathbb R}$, for every $\epsilon>0$, there exists  $\delta>0$ such that 
$$||f-g||_{\infty}\leq \delta\Rightarrow |H(f)-H(g)|
\leq \epsilon.$$


\noindent{\bf Proof.} 
We first need to introduce some definitions and notations. 
\\
For a finite set of points $S=\{a_1,\dots,a_n\}$ in $\mathbb{R}^2$:  $|S|=n$;
$a_i=(x_i,y_i)$, $\ell_i=y_i-x_i$ and $p_i=\frac{\ell_i}{L}$, for all $i$, being  $L=\sum_{i=1}^n \ell_i$.
\\
 For points $a = (x,y)$ and $a' = (x', y')$: 
$||a-a'||_{\infty}=\max\{|x-x'|,|y-y'|\}$.
Similarly, $||f-g||_{\infty}=\sup\{|f(q)-g(q)|: q\in\mathbb {R}^n\}$.
\\
The bottleneck distance
between the persistence diagrams $D(f)$ and $D(g)$ associated  to the lower star filtrations of $f$ and $g$ is:
$d_B(D(f),D(g))=\inf_{\gamma}\sup_{a}\{||a-\gamma(a)||_{\infty}\}$,
where bijection $\gamma:D(f)\to D(g)$  can associate a point off the diagonal with another on
the diagonal or both off the diagonal\footnote{{\it Diagonal} is the set of points $\{(x,x)\}\subset \mathbb{R}^2$.}. 
\\
Now, let $n=\max\{|D(f)|, |D(g)|\}$. 
Let  $\gamma: D(f)\to D(g)$ be the bijection such that  $d_B(D(f),D(g))=\sup_{a}\{||a-\gamma(a)||_{\infty}\}$.
Then $D(f)=\{a_1,\dots,a_n\}$ and  $D(g)=\{a'_1,\dots,a'_n\}$,
being $a'_i=\gamma(a_i)$ for all $i$.
\\
Since $h(x)=-x\log x$ is a continuous function in $[0,1]$ (redifining $h(0)$ as $0$), for $\epsilon'=\frac{\epsilon}{n}>0$, there exists $\delta'\in (0,1]$ such that if $|x-x'|\leq \delta'$ then
$|h(x)-h'(x)|\leq \epsilon'$.
\\
Stability of persistence diagram \cite[page 105]{stability} establishes that $d_B(D(f), D(g))$ $\leq ||f -g||_{\infty}$.
 So if $||f -g||_{\infty}\leq \delta$, then
$|\ell_i-\ell'_i|\leq   
2\delta$
for all $i$ and $|L-L'|\leq
2\delta n$.
\\
Without loss of generality, assume that $L\geq L'$.
 Let $\delta=\frac{\delta' L'}{4n}> 0$.
Then, $|p_i-p'_i|\leq
\delta'$
for all $i$:
\begin{itemize}
\item $p_i-p'_i
=\frac{\ell_i}{L}-\frac{\ell'_i}{L'}
=\frac{L'\ell_i-L\ell'_i}{LL'}
\leq \frac{\ell_i-\ell'_i}{L'}
\leq \frac{2\delta}{L'}
=\frac{2\delta' L'}{4nL'}
=\frac{\delta'}{2n}\leq \delta'$.
\item $p'_i-p_i
\leq\frac{(L'+2\delta n)\ell'_i-L'\ell_i}{LL'}
\leq \frac{2\delta}{L}+\frac{2\delta n\ell'_i}{LL'}
%
= \frac{2\delta' L'}{4nL}+\frac{2\delta' L'n\ell'_i}{4nLL'}
\leq \frac{\delta'}{2n}+\frac{\delta' }{2}\leq \delta'$.
\end{itemize}
Therefore, $|H(f)-H(g)|=
|\sum_{i=1}^n p_i\log p_i-\sum_{i=1}^np'_i\log p'_i|\leq
\sum_{i=1}^n |p_i\log p_i-p'_i\log p'_i|\leq n\epsilon'=\epsilon$
which concludes the proof.
\hfill{$\Box$}

Now, suppose we have two discrete signals $\bar{f}:\mathbb{R}^n\to c\mathbb{Z}$ and
$\bar{g}:\mathbb{R}^n\to 
c\mathbb{Z}$,  being $c$ a small positive number (machine precision).
Suppose that $D(g)$ has $n$ points. 
Since $h(x)=-x\log x\in [0, \frac{1}{e}]$, then 
$\epsilon'\in (0, \frac{1}{e}]$ and $\epsilon\in (0,\frac{n}{e}]$.
If, for example,  $n=1000$, then    $\epsilon\in (0, 368]$.
Write  $L'=n\ell'$, being $\ell'$ the average length of all the intervals in $D(g)$. 
Then $\delta'=\frac{4n\delta}{L'}=\frac{4\delta}{ \ell'}$. To have $0<\delta'<<1$, we need 
$\ell'$ to be large with respect to $\delta$. For example, for $\delta'\leq 0.1$,   we need  $\ell'\geq 40\delta$, which is equivalent to say that the average between consecutive local max and local min in $g$ should be at least  $40$ times the difference between $f$ and $g$, and, in particular, at least $40$ times the machine precision $c$. 
In practice, this is a common case.

Consider, for example, the two signals $f$ and $f'$ in Figure \ref{fig:dcomotors}. In this signals,
the machine precision is $c= 1.4386\cdot 10^{-14}$.
The number of points in the respective persistent diagrams are: $|D(f)|=421$ and $|D(f')|=7213$, being the total lengths: $L=488.5810$ and $L'= 9582.5$.
Averge lengths are: $\ell=\frac{L_1}{|D(f)|}=1.1605$ and $\ell'=
\frac{L'}{|D(f')|}=1.3285$.
Observe that $\ell$ is $0.8\cdot 10^{14}$ times $c$ and $\ell'$ is $0.9\cdot 10^{14}$ times $c$.
Finally, $H(f)=5.9977$ and $H(f')=8.7997$.

For practical scopes, the methodology explained above can be translated in the following algorithm
designed for analyzing a $2-$dimensional plot.

Suppose that the first coordinate of a point in $\mathbb{R}^2$ represents time. 
Given a signal $S\subset \mathbb{R}^2$:
\begin{itemize}
\item order the points in $S$ respect to their first coordinate (i.e., order the points in $S$ by time);
\item transform $S$ into a filtered simplicial complex: 
\begin{itemize}
\item 
each point of $S$ is a $0-$simplex with filter equal to 
its second coordinate.
\item Each pair formed by two consecutive points in $S$: $ (x_i,y_i)$ and $(x_{i+1},y_{i+1}) \in S$ where $x_i<x_{i+1}$, forms a $1-$simplex $\sigma$ with filter value $f(\sigma)=max\{y_i,y_{i+1}\}$. Note that 
the resulting filration is obtained by presenting at the beginning the simplices formed with the lowest second coordinate (i.e., the  filter-value set $F$ is obtained by spanning the $Y-$axis in a upward direction).
\end{itemize}
The resulting filtration is a {\it lower start filtration}.
\item compute \textit{persistent entropy}.
\end{itemize}
Figure ~\ref{fig:construction} shows an example of the application of the methodology. From left to right: a) The input signal formed by three time points, respectively with coordinates: $(1,0),~(2,2),~(3,1)$. b) The filtered simplicial complex formed by three $0-$simplices: $\{v_0,v_1,v_2\}$ with filter values $f(v_0)=0, f(v_2)=1, f(v_1)=2$ and two $1-$simplices: $\{e_1, e_2\}$, with filter values $f(e_1)=f(e_2)=2$, so the filter-value set is $F=\{0,1,2\}$. c) The persistent barcodes: at $F=0$, there is only one topological feature corresponding to $v_0$; at $F=1$, $v_0$ is still in the space but also a new component is introduced and it corresponds to $v_2$; eventually for $F=2$, a new $0-$simplex is added to the topological space $(v_1)$ within the two $1-$simplices $e_1,~e_2$ where $e_1=\{v_0,v_2\}$ and $e_2=\{v_2,v_1\}$. From this filter value and successive, the space is described by only one persistent connected component, i.e. $\beta_0=1$. Visually there is only \textit{one infinite line in the barcode}. \\
The persistent entropy of the space is computed as follows. The maximum filer value is $2$, so the symbol ``$\infty$'' representing the persistent topological feature is substituted with the value $m=3$. Then, the barcode is formed by two lines with lengths $\ell_1=1$ and $\ell_2=3$, respectively. So the total length $L=1+3=4$, for each line the probability is given by $p_1=1/4$ and $p_2=3/4$, and finally the persistent entropy is $H=0.5623$.

\subsection{The case study: comparison of DC motors}
\label{sec:results}
\begin{figure}
\centering
\includegraphics[width=0.8\textwidth]{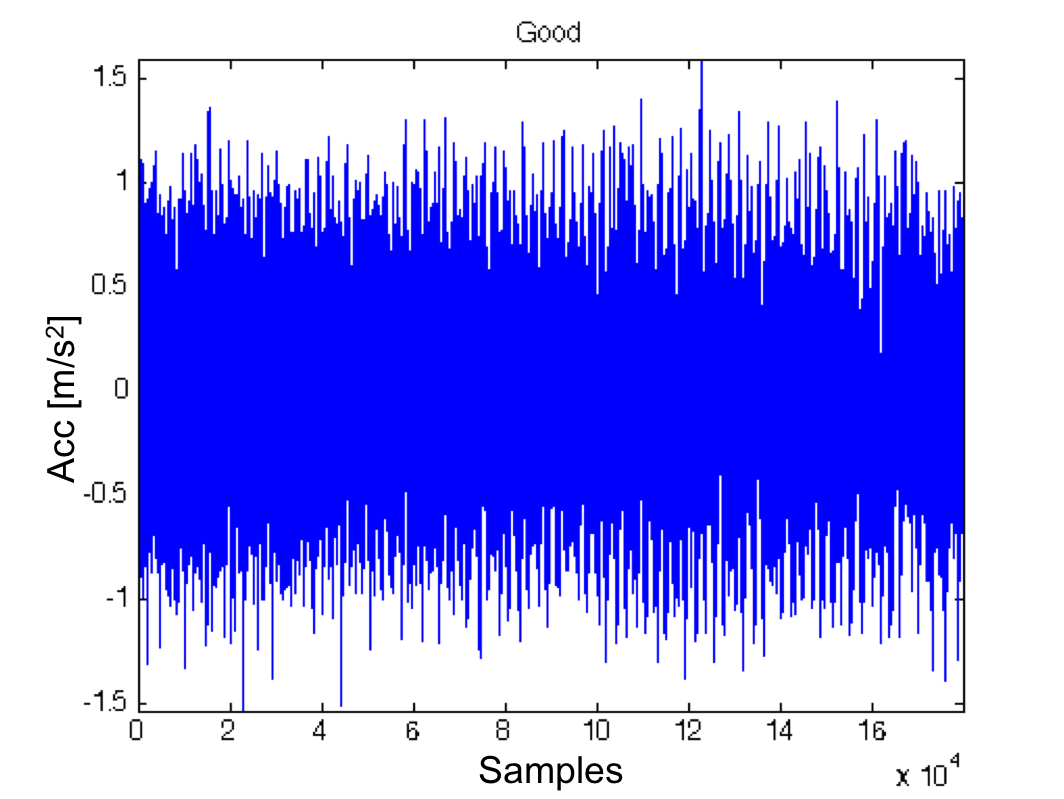}
\includegraphics[width=0.8\textwidth]{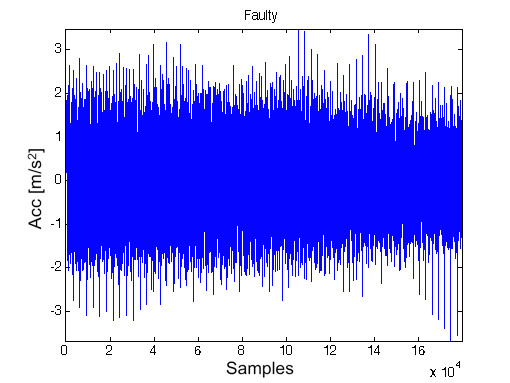}
\caption{Example of signals of two DC motors
with 180,000 time points and sampling frequency 50kHz. 
Top: Signal $f_1$ of a {\it good} motor. Bottom: Signal 
$f_2$ of a {\it faulty} motor.
The motors were classified by an expert operator based on their vibration and noise level.
}
\label{fig:dcomotors}
\end{figure}

We applied our methodology to 46 small DC motors (see two examples in Figure~\ref{fig:dcomotors}). 
For each motor we analyzed the acceleration that has been measured with a B\&K single axis 4514-001 IEPE accelerometer for acquiring the radial component of vibration. Signals were sampled at a rate of 50 kHz with a total number of 180,000 time points. For all the detailed information regarding the DC motors and data acquisition we refer to~\cite{concettoni2012mechanical,rucco2015topological}.

For the sake of preciseness, in this work we used a subset of the set reported in~\cite{concettoni2012mechanical,rucco2015topological} that is formed by signals with the same length.
 Each signal is formed by 180000 number of points equally time-spaced.
The number of simplices (vertices and edges) for each signal is  359999  (the signals have the same number of vertices and edges but the length of the edges is different). The machine precision (minimum value between two signal values) is: $1.4386\cdot 10^{-14}$.

\begin{figure}
\centering
\includegraphics[width=\textwidth]{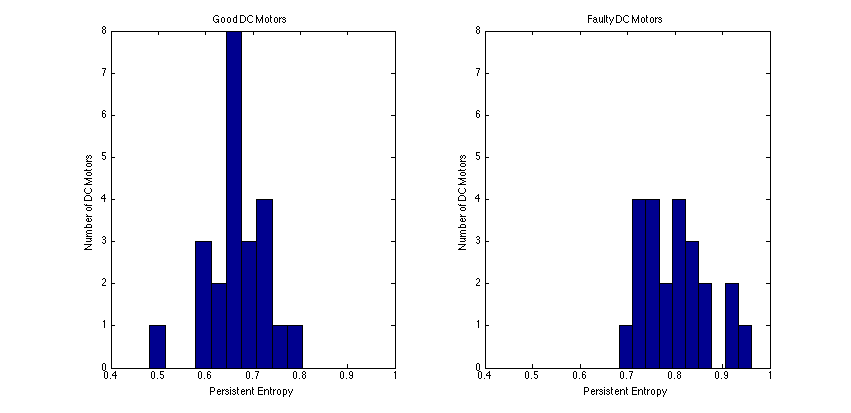}\\
\caption{Histogram of the persistent entropy. Left: distribution for \textit{good motors} with average value $\bar{H}=0.6647$ and standard deviation 0.0668. Right:  distribution for \textit{good motors} with average value $\bar{H}=0.8001$ and standard deviation 0.0711}
\label{fig:histogram}
\end{figure}

\begin{figure}
\centering
\includegraphics[width=\textwidth]{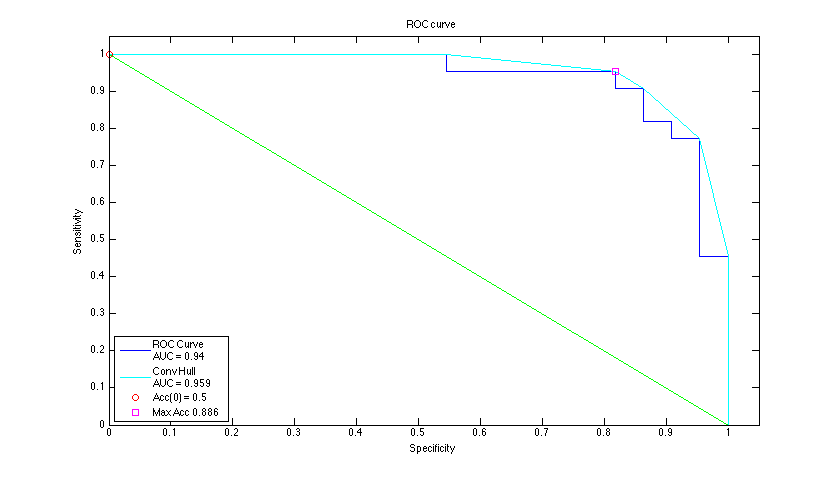}\\
\caption{ROC curve of the classifier based on  persistent entropy. From the analysis of the ROC curve we derived the threshold $\theta=0.7211$ that maximizes the accuracy of classification. Where accuracy means the ratio between the {\it number of correct assessments and the number of all assessments}~\cite{ROC}.}
\label{fig:roc}
\end{figure}

\begin{figure}
\centering
\includegraphics[width=\textwidth]{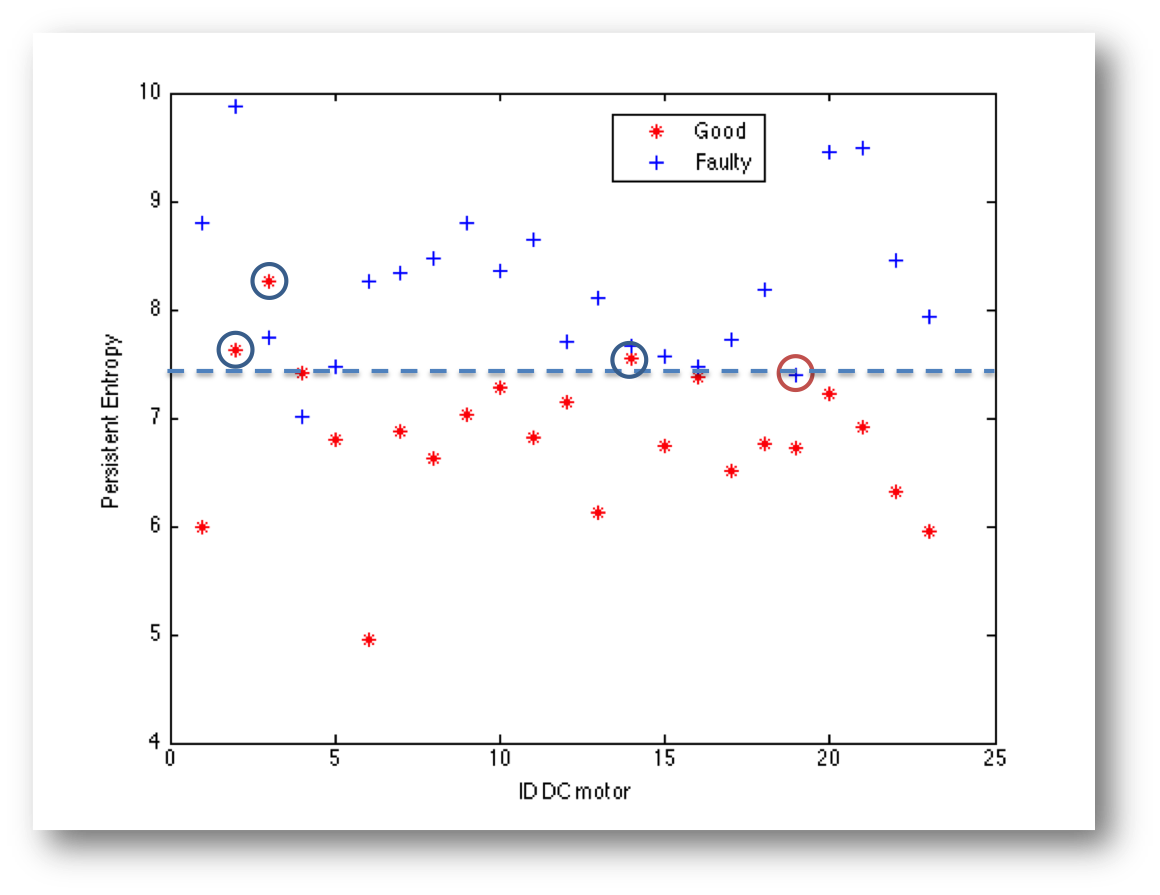}\\
\caption{Plot of the persistent entropy H for each motor. All the good motors have less persistent entropy than the persistent entropy of the faulty motors. The dashed lines is the separation line with constant value  $\theta=0.7211$, the threshold that maximizes the accuracy of the ROC curve. 
}
\label{fig:separation}
\end{figure}

The software has been coded in MATLAB and for the topological analysis we used the Java package Javaplex~\cite{adams2011javaplex}.  The motors were classified by an expert operator in two classes: \textit{good motors, and faulty motors} based on their vibration and noise level. The persistent entropy is used for defining a 1-dimensional feature space that is used for classifying the motors (see Figure~\ref{fig:histogram} and Figure \ref{fig:separation}) The quality of this feature is evaluated by a ROC curve (see Figure~\ref{fig:roc}) with a k-cross validation (k=7). The Area Under Curve (AUC) is AUC=94.52\%. The threshold that maximizes the \textit{accuracy} is $\theta=0.7211$ that can be used for classifying future motors (dashed lines in the Figure~\ref{fig:separation}).

The average computational time for each motor is in the order of 180 seconds on the following laptop:
MacBook Air, 1.7GHz i5, GB RAM, Hard Disk SSD.

\section{Discussion}
\label{sec:conclusions}
In this paper we reported on the definition of a new technique, based on persistent homology and information theory, for comparing discrete signals. The transformation of a signal into a filtered simplicial complex of dimension 1 lets to study its topology in terms of persistent homology. The persistent barcodes characterize the signals and they are used for calculating an entropy measures, the so-called persistent entropy. Persistent entropy is used as feature for comparing signals. One of the main outcome of this paper is the {\it stability theorem for the persistent entropy}. This theorem gives the formal support for the comparison of the persistent entropy of two signals.
The methodology presented in this paper is able to state if two signals have the same shape even if one is shifted respect to the other. More rigorously, given two signals $(S_1,S_2)$ with the same shape but $S_2$ is shifted respect to $S_1$ $(x_2=x_1+n~and/or~y_2=y_1+m$,where $m,n \in \mathbb{R})$, the persistent entropy for $S_1$ and $S_2$ is the same. The y-shifting increases the filter values $(F2=F1+m)$ but the number of lines within a barcode and their lengths is completely preserved. 
In future investigation we will verify if our methodology is more computational convenient respect to the computation of distances usually used for measuring similarities between two persistent barcode (e.g., Wasserstein and Bottleneck). We also plan to use the methodology for comparing the time-varying persistent entropy plots used for describing the persistent entropy automaton (PEA) in the $S[B]$ approach~\cite{merelli2015topological}, eventually we intend to deal with higher dimensional multivariate signals.

\section*{Acknowledgments}
We acknowledge the financial support of the Future and Emerging Technologies (FET) programme within the Seventh Framework Programme (FP7) for Research of the European Commission, under the FP7 FET-Proactive Call 8 - DyMCS, Grant Agreement TOPDRIM, number FP7-ICT-318121 and Science and Innovation Spanish Ministry under project number MTM2012-32706.
\bibliographystyle{elsarticle-num} 
\bibliography{typeinst}
\end{document}